\documentclass{elsart}
\usepackage{amsfonts}
\usepackage{amsmath}
\usepackage{amssymb}
\usepackage{graphicx}

\begin{document}

\begin{frontmatter}
\title{Analytical calculation of the solid angle subtended by a circular disc detector
at a point cosine source\thanksref{FCT}}
\thanks[FCT]{Partially supported by Funda\c{c}\~{a}o para a Ci\^{e}ncia e Tecnologia
(Programa Praxis XXI - BD/15808/98)}
\author{M. J. Prata\thanksref{TelFax}}
\thanks[TelFax]{Tel.: +351-21-944-0690; fax: +351-21-846-3276.}
\ead{mjprata@sapo.pt}
\address{Instituto  Tecnol\'{o}gico e Nuclear (ITN), Estrada Nacional 10, Sacav\'{e}m 2686-953, Portugal}
\begin{abstract}
We derive analytical expressions for the solid angle subtended by a
circular disc at a point source with cosine angular distribution ($f(\mu)=\mu/\pi$)
under the sole condition that the disc lies in the half-space illuminated by the source ($\mu \geq 0$). The expressions are given with reference to two alternative coordinate systems (S and S'), S being such that the $z$ axis is parallel to the symmetry axis of the disc and
S' such that the $z'$ axis is aligned with the source direction. Sample plots of the expressions are presented.
\end{abstract}
\end{frontmatter}

\section{Introduction}

In two recent works we obtained analytical expressions for the solid angle
subtended by a cylindrical shaped detector at a point cosine source in the
cases where the source axis is orthogonal \cite{Prata2003a} or parallel \cite%
{Prata2003c} to the cylinder axis of revolution. As ancillary results we
also derived expressions for the solid angle defined by a circular disc in
the cases where the source axis is orthogonal ($\Omega_{\perp}$) or parallel
($\Omega_{\parallel}$) to the symmetry axis of the disc. This latter result (%
$\Omega_{\parallel}$) appeared in a previous work by Hubbell \textit{et al} 
\cite[eq. 29]{Hubb61}, where it is also credited to other authors \cite%
{Herm00,Foot15}. In that work, a quite general treatment of the radiation
field due to a circular disc source with axial symmetry is given in terms of
a Legendre expansion and $\Omega_{\parallel}$ appears as a subsidiary result
which is interpreted as the response of a plane detector parallel to a
Lambertian uniformly distributed disc source. Circular apertures and sources
are often considered in optics and radiation physics; and disc-shaped
detectors are widely used in nuclear science (e.g. Neutron Activation
Analysis). References to other works on this subject in the context of
nuclear physics can be found in \cite{Knol79} and \cite{Tsou95}. While the
case of a point isotropic source has been treated to great extent \cite%
{Jaff54,Mack56,Mask56,Mask57,Gard71,Prata2003b}, the case of a point cosine
source has, to the best of our knowledge, attracted little attention, with
the already mentioned exception of \cite{Hubb61}. For these reasons, in the
present work we extend the scope of the previous results by performing the
calculation of the solid angle defined by a circular disc and a point cosine
source pointing at an arbitrary direction, under the sole restriction that
the disc lies in the half-space illuminated by the source.

The solid angle subtended by a given surface at a point source located at
the origin can be defined by

\begin{equation}
\Omega_{surf}=\iint\limits_{\substack{ directions  \\ hitting~surface}}f(%
\mathbf{\Omega})d\Omega~,  \label{eq_omega_surf1}
\end{equation}

where $f(\mathbf{\Omega})d\Omega$ is the source angular distribution. In the
case of a point cosine the distribution is defined in relation to some
direction axis specified by the unit vector $\mathbf{k}$ and it is given by $%
f(\mathbf{\Omega})=(\mathbf{\Omega\cdot k+}\left\vert \mathbf{\Omega\cdot k}%
\right\vert )/(2\pi)$, the $(2\pi)^{-1}$ factor ensuring that $0\leq
\Omega_{surf}\leq1$. For $\mu=\mathbf{\Omega\cdot k}$ it follows that $f(%
\mathbf{\Omega})=\{\mu/\pi~(\mu\geq0);0~(\mu<0)\}$ so that the source only
emits into the hemisphere around $\mathbf{k}$. Because of this, the
integration limits in the RHS of eq. \ref{eq_omega_surf1} are to be
determined from the conditions that $\mu\geq0$ and that each included $%
\mathbf{\Omega}$ direction hits the surface. In the following we shall
assume that the position of the disc is always chosen is such a manner that $%
\mu\geq0$ for each point on the disc. This restriction greatly simplifies
the calculation without, we believe, reducing the practical interest of the
expressions. This is so because, in actual situations, the source is
distributed over some planar surface; the source axis is coincident with the
normal to the surface and then the restriction simply requires that the
detector be held somewhere on the illuminated side of the radiating surface
or eventually directly on the surface but does not intersect it.

\section{Solid Angle Calculation\label{section_solid_angle}}

\begin{figure}[h]
\begin{center}
\includegraphics[
height=7.4422cm,
width=6.0934cm
]{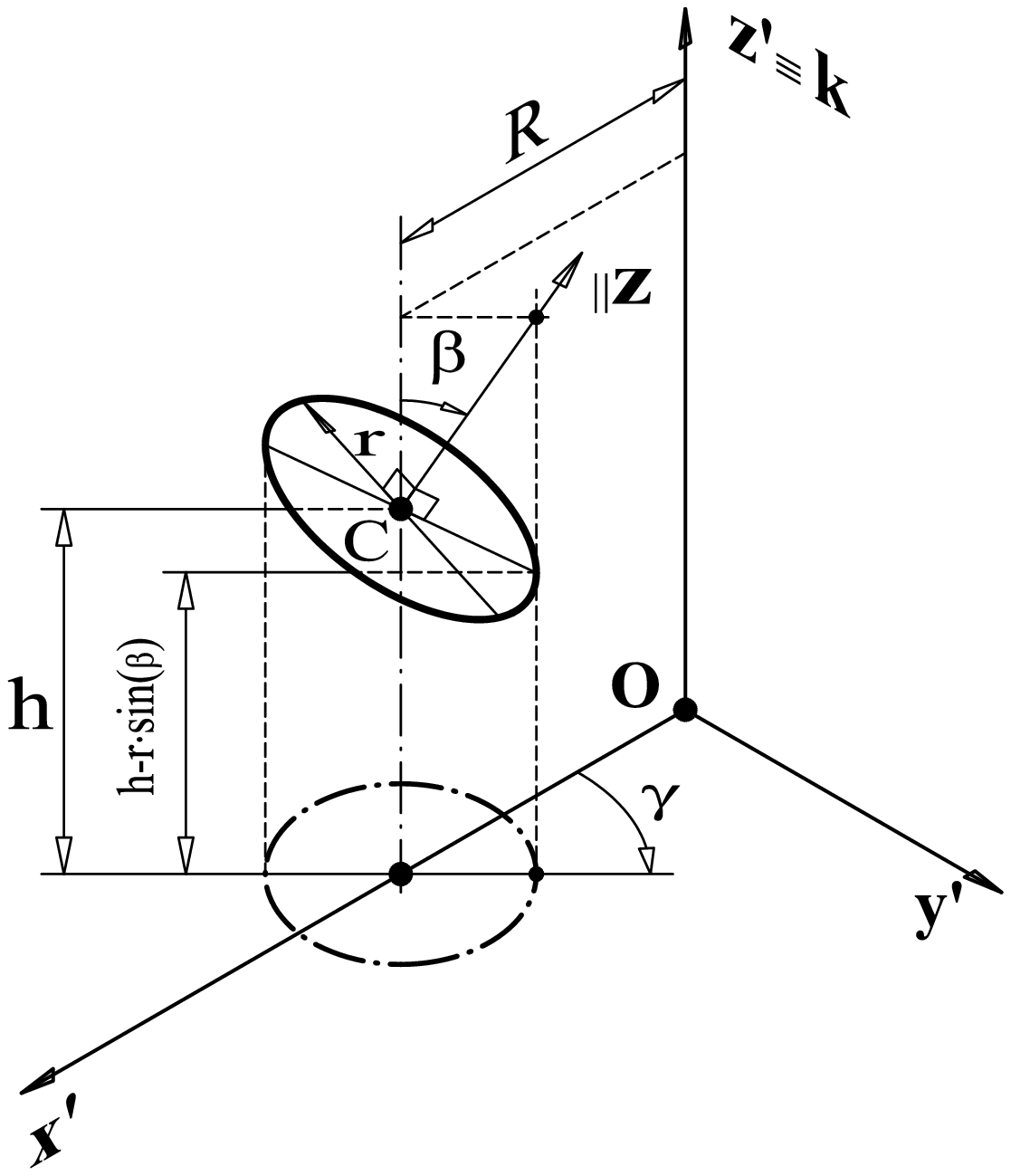}
\end{center}
\caption{Notation used in the S' coordinate system.}
\label{fig1}
\end{figure}

\begin{figure}[h]
\begin{center}
\includegraphics[
height=5.1687cm,
width=6.0934cm
]{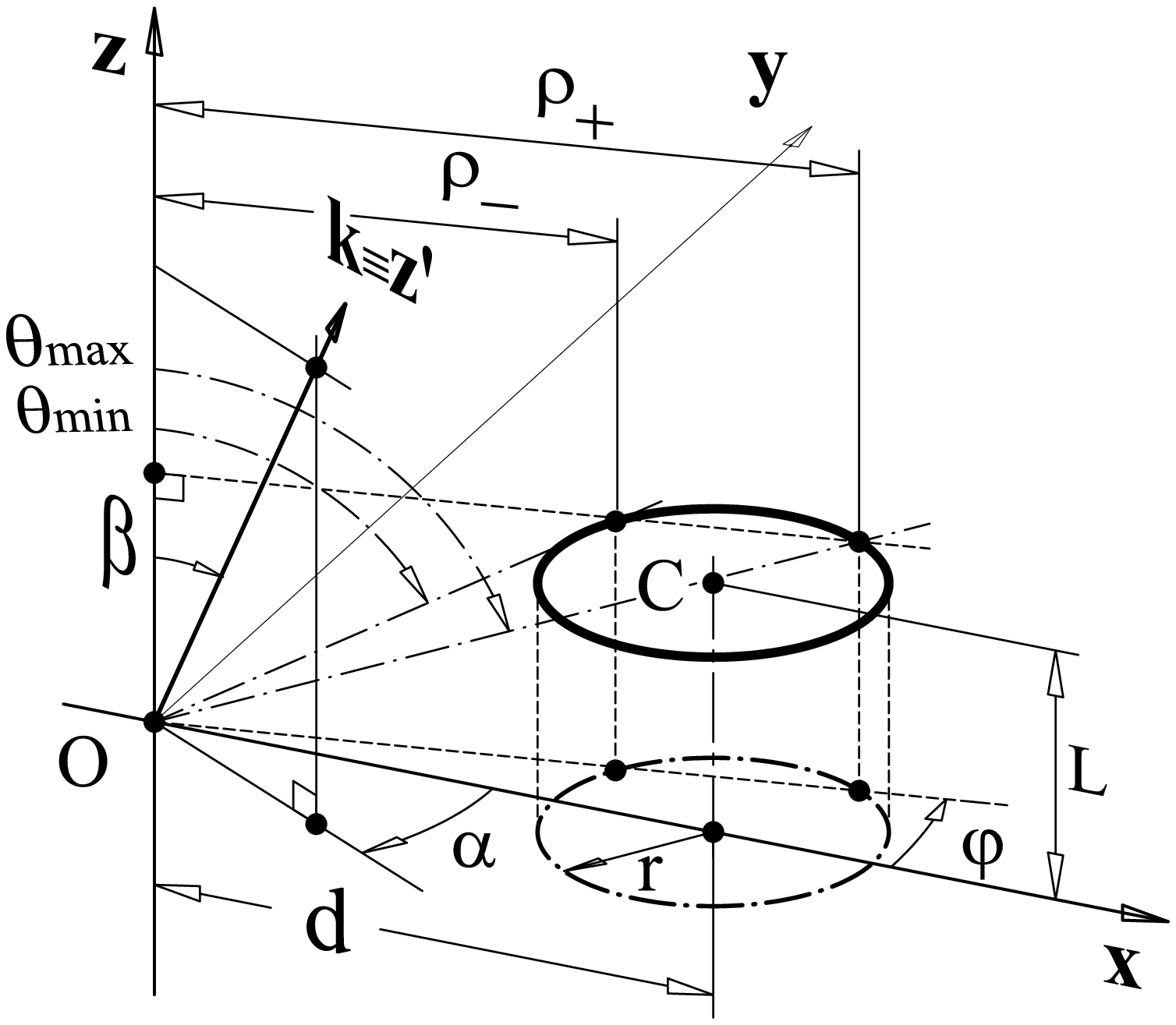}
\end{center}
\caption{Notation used in the S coordinate system.}
\label{fig2}
\end{figure}

To proceed it is advantageous to consider two coordinate systems (S' and S)
with a common origin also coincident with the source position. In the S'
system (fig. \ref{fig1}) the $z^{\prime }$ axis is aligned with the source
direction $\mathbf{k}$; the position of the disc center (C) is specified by $%
h$ and $R$; and the symmetry axis of the disc is specified by the angles $%
\beta $ and $\gamma $, $\beta $ being the angle between the source and the
disc axes. When working in the S system (fig. \ref{fig2}), the $z$ axis is
parallel to the disc axis, C is located by means of $L$ and $d$; and $%
\mathbf{k}$ is given by angles $\beta $ and $\alpha $. Due to the symmetry
of the source it is possible to choose the $x,y$ axes so that the $y$
coordinate of C is zero in each coordinate system. Also from the symmetry of
the source it is clear that the solid angle is an even function of $\alpha $
or $\gamma $ and in the following calculations we will thus assume that $%
0\leq \alpha \leq \pi $ and $0\leq \gamma \leq \pi $. Finally, it is
sufficient to consider $0\leq \beta \leq \pi /2$.

The restriction that $\mu\geq0$ over the whole disc is easily expressed in
S' by 
\begin{equation}
h-r\cdot\sin\beta\geq0~,  \label{eq_hbigger_r}
\end{equation}
$r>0$ being the disc radius. This is automatically satisfied, provided $%
h\geq r$, but otherwise, it further reduces the range of variation of $\beta$
to

\begin{equation}
0\leq\beta\leq\arcsin(h/r)~,h<r~.  \label{eq_beta_range_h_less_r}
\end{equation}

It is readily found that the coordinates of a point in S' and S ($\mathbf{P}%
^{\prime}$ and $\mathbf{P}$) are related by an orthogonal matrix ($\mathbf{M}
$):

\begin{equation}
\mathbf{P}^{\prime}=\mathbf{M\cdot P}~,  \label{eq_Plin_M_P}
\end{equation}

\begin{equation}
\mathbf{P}=\mathbf{M}^{\mathbf{T}}\mathbf{\cdot P}^{\prime}~,
\label{eq_P_Mt_Plin}
\end{equation}

where $\mathbf{M}$ is given by 
{\footnotesize 
\begin{equation}
\mathbf{M}(\gamma,\beta,\alpha)=%
\begin{bmatrix}
\sin{\gamma}\sin{\alpha+}\cos{\gamma}\cos{\beta}\cos{\alpha~~} & \sin{\gamma 
}\cos{\alpha-}\cos{\gamma}\cos{\beta}\sin{\alpha~~} & {-}\cos{\gamma}\sin{%
\beta} \\ 
\cos{\gamma}\sin{\alpha-}\sin{\gamma}\cos{\beta}\cos{\alpha~~} & \cos{\gamma 
}\cos{\alpha+}\sin{\gamma}\cos{\beta}\sin{\alpha~~} & \sin{\gamma}\sin{\beta 
} \\ 
\sin{\beta}\cos{\alpha} & {-}\sin{\beta}\sin{\alpha} & \cos{\beta}%
\end{bmatrix}
\label{eq_M_def}
\end{equation}
}

and its inverse is given by the transpose {\footnotesize 
\begin{equation}
\mathbf{M}^{\mathbf{T}}(\gamma,\beta,\alpha)\mathbf{=}%
\begin{bmatrix}
\sin{\gamma}\sin{\alpha+}\cos{\gamma}\cos{\beta}\cos{\alpha~~} & \cos{\gamma 
}\sin{\alpha-}\sin{\gamma}\cos{\beta}\cos{\alpha~~} & \sin{\beta}\cos{\alpha 
} \\ 
\sin{\gamma}\cos{\alpha-}\cos{\gamma}\cos{\beta}\sin{\alpha~~} & \cos{\gamma 
}\cos{\alpha+\sin\gamma}\cos{\beta}\sin{\alpha~~} & {-}\sin{\beta}\sin{%
\alpha } \\ 
{-}\cos{\gamma}\sin{\beta} & \sin{\gamma}\sin{\beta} & \cos{\beta}%
\end{bmatrix}
~.  \label{eq_Mt_def}
\end{equation}
}

Let $(x_{c},y_{c},z_{c})$ and $(x_{c}^{\prime },y_{c}^{\prime
},z_{c}^{\prime })$ denote the coordinates of the disk center in S and S',
respectively. As previously said, it is possible to choose the $x,y$ axes in
each coordinate system so that $y_{c}=y_{c}^{\prime }=0$. Starting in the S
system, setting $(x_{c},y_{c},z_{c})=(d,0,L)$ and fixing $\alpha $ and $%
\beta $, one then uses eq. \ref{eq_Plin_M_P} to obtain $(x_{c}^{\prime
},y_{c}^{\prime },z_{c}^{\prime })$. The value of $\gamma $ is determined by
imposing that $y_{c}^{\prime }=0$, which gives

\begin{equation}
\tan\gamma=(d\sin\alpha)/(d\cos\beta\cos\alpha-L\sin\beta)~,
\label{eq_gamma_star}
\end{equation}
or, for $0\leq\gamma\leq\pi$,

\begin{gather}
\gamma =\gamma (\beta ,L,d,\alpha )  \notag \\
=\arccos [(d\cos \beta \cos \alpha -L\sin \beta )/\sqrt{(d\sin \alpha
)^{2}+(d\cos \beta \cos \alpha -L\sin \beta )^{2}}~]~.
\label{eq_gamma_arccos}
\end{gather}%
The values of $h$ and $R$ are obtained from

\begin{equation}
h=z_{c}^{\prime}=h(\beta,L,d,\alpha)=d\sin\beta\cos\alpha+L\cos\beta~,
\label{eq_h_Ld}
\end{equation}
and

\begin{align}
R& =x_{c}^{\prime }=\sqrt{x_{c}^{\prime 2}+y_{c}^{\prime 2}}  \notag \\
& =R(\beta ,L,d,\alpha )  \notag \\
& =\sqrt{d^{2}(\sin ^{2}\alpha +\cos ^{2}\alpha \cos ^{2}\beta )+L^{2}\sin
^{2}\beta -dL\sin (2\beta )\cos \alpha }~.  \label{eq_R_Ld}
\end{align}%
Eq. \ref{eq_gamma_arccos} can then be written as

\begin{equation}
\gamma =\arccos [(d\cos \beta \cos \alpha -L\sin \beta )/R]~.
\label{eq_gamma_R}
\end{equation}

Conversely, starting in the S' system with $(x_{c}^{\prime },y_{c}^{\prime
},z_{c}^{\prime })=(R,0,h)$, using eq. \ref{eq_P_Mt_Plin} to obtain $%
(x_{c},y_{c},z_{c})$ and imposing that $y_{c}=0$, there results

\begin{gather}
\alpha =\alpha (\beta ,h,R,\gamma )  \notag \\
=\arccos [(R\cos \beta \cos \gamma +h\sin \beta )/\sqrt{(R\sin \gamma
)^{2}+(R\cos \beta \cos \gamma +h\sin \beta )^{2}}~]~.
\label{eq_alfa_arccos}
\end{gather}

\begin{equation}
L=z_{c}=L(\beta,h,R,\gamma)=h\cos\beta-R\sin\beta\cos\gamma~,
\label{eq_L_hr}
\end{equation}
and

\begin{align}
d& =x_{c}=\sqrt{x_{c}^{2}+y_{c}^{2}}  \notag \\
& =d(\beta ,h,R,\gamma )  \notag \\
& =\sqrt{R^{2}(\sin ^{2}\gamma +\cos ^{2}\gamma \cos ^{2}\beta )+h^{2}\sin
^{2}\beta +Rh\sin (2\beta )\cos \gamma }~.  \label{eq_d_hr}
\end{align}%
Again, eq. \ref{eq_alfa_arccos} \ can be cast as

\begin{equation}
\alpha =\arccos [(R\cos \beta \cos \gamma +h\sin \beta )/d]~.
\label{eq_alfa_d}
\end{equation}

The substitution of the expression for $h$ (eq. \ref{eq_h_Ld}) on the RHS of
eq. \ref{eq_hbigger_r} and a bit of algebra gives the restriction $\mu \geq
0 $ expressed in terms of S variables:

\begin{equation}
L\geq(r-d\cos\alpha)\cdot\tan\beta~.  \label{eq_Lbigger_r}
\end{equation}
If $\beta=\pi/2$, the previous eq. can be rewritten as 
\begin{equation}
\cos\alpha\geq r/d~.
\end{equation}

The solid angle (eq. \ref{eq_omega_surf1}) is best calculated in S. With the
notation described in fig. \ref{fig2}, it is seen that $k=(\sin\beta\cos
\alpha,-\sin\beta\sin\alpha,\cos\beta)$ and

\begin{equation}
\mathbf{k\cdot\Omega=}\cos\beta\cos\theta+\sin\beta\sin\theta\cos
(\alpha+\varphi)~,
\end{equation}
where, for a given direction, $\theta$ is the polar angle from the $z$ axis
and $\varphi$ is the azimuthal angle in the $xy$ plane measured from the $x$
axis. The solid angle is then given by

\begin{equation}
\Omega=\pi^{-1}\textstyle\int\nolimits_{\varphi_{\min}}^{\varphi_{\max}}%
\textstyle\int\nolimits_{\theta_{\min}}^{\theta_{\max}}k\cdot\Omega
\sin(\theta)d\theta d\varphi~,  \label{eq_omega_1}
\end{equation}
Because the position of the disc is such that $\mu\geq0$ (eq. \ref%
{eq_Lbigger_r} or, equivalently, eq. \ref{eq_hbigger_r}), the integrations
limits are determined only by the condition that each included $(\theta
,\varphi)$ direction hits the detector.

\begin{figure}[h]
\begin{center}
\includegraphics[
height=4.323cm,
width=6.0912cm
]{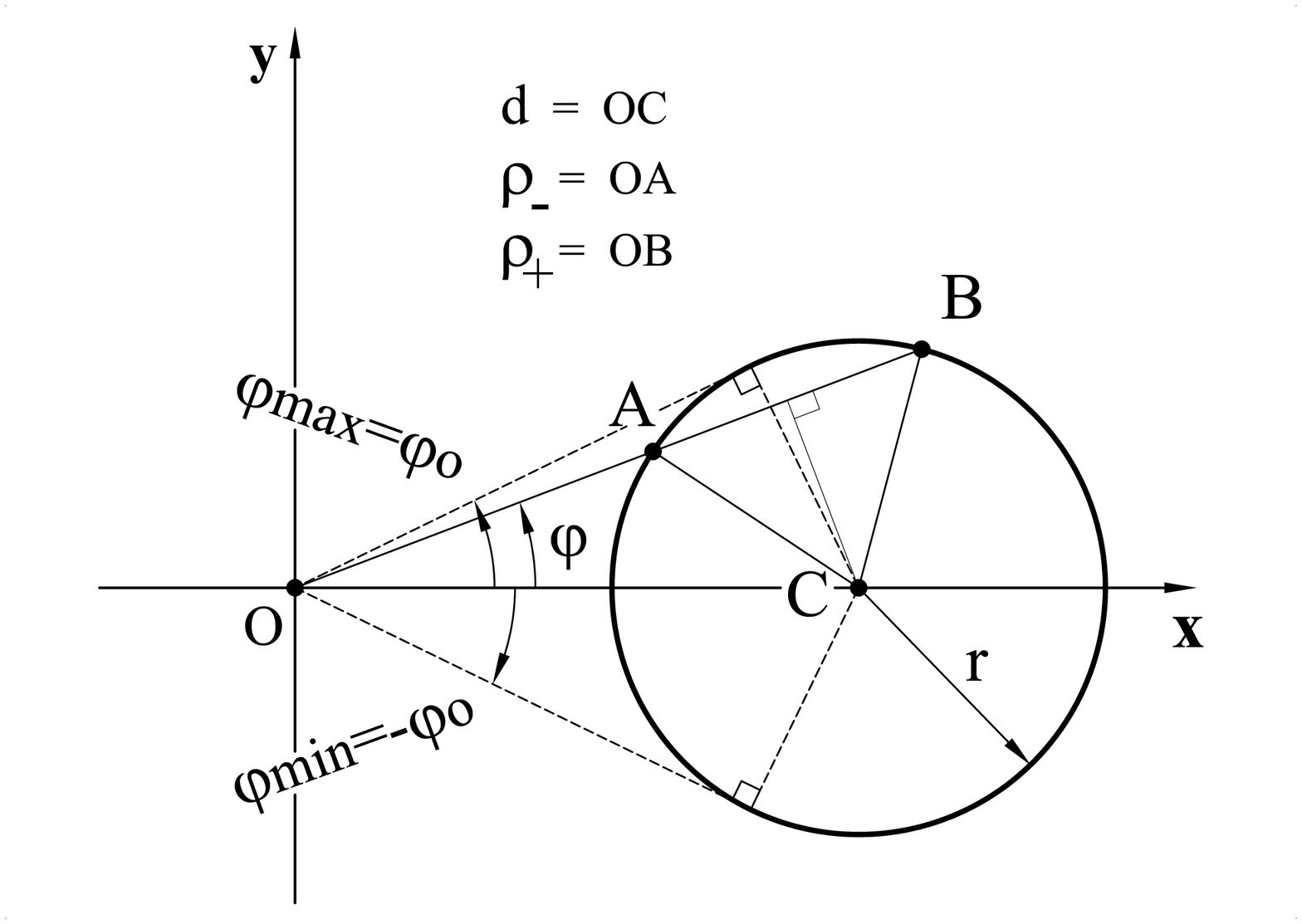}
\end{center}
\caption{Integrations limits for $\protect\varphi$ and definitions of $%
\protect\rho_{\pm}$.}
\label{fig3}
\end{figure}
Referring to figs. \ref{fig2} and \ref{fig3}, it follows that $%
\varphi_{\max}=-\varphi_{\min}=\varphi_{o}\equiv\arcsin(r/d)~,$

$\theta_{\max}=\arccos(L/\sqrt{L^{2}+\rho_{+}^{2}})~,$

$\theta_{\min}=\arccos(L/\sqrt{L^{2}+\rho_{-}^{2}})~,$

where

$\rho_{\pm}=d\cos\varphi\pm\sqrt{r^{2}-(d\sin\varphi)^{2}}~.$

Then,

\begin{align}
\Omega & =\pi ^{-1}(\cos \beta {\int\nolimits_{-\varphi o}^{+\varphi o}}{%
\int\nolimits_{\theta _{\min }}^{\theta _{\max }}}\cos \theta \sin \theta
~d\theta d\varphi +  \notag \\
& \sin \beta {\int\nolimits_{-\varphi o}^{+\varphi o}}\cos (\alpha +\varphi )%
{\int\nolimits_{\theta _{\min }}^{\theta _{\max }}}\sin ^{2}\theta ~d\theta
d\varphi )~,  \label{eq_omega_2} \\
& =\cos \beta ~\Omega _{\parallel }+\sin \beta ~\Omega _{\perp }~,
\label{eq_omega_3}
\end{align}%
where

\begin{equation}
\Omega_{\parallel}=\pi^{-1}\int\nolimits_{-\varphi o}^{+\varphi
o}\int\nolimits_{\theta_{\min}}^{\theta_{\max}}\cos\theta\sin\theta~d\theta
d\varphi~
\end{equation}
and

\begin{equation}
\Omega_{\perp}=\pi^{-1}\int\nolimits_{-\varphi o}^{+\varphi o}\cos
(\alpha+\varphi)\int\nolimits_{\theta_{\min}}^{\theta_{\max}}\sin^{2}%
\theta~d\theta d\varphi.
\end{equation}
The values of $\Omega_{\parallel}$ and $\Omega_{\perp}$were previously
obtained. From \cite[eq. 29]{Hubb61}, or \cite[eq. 33]{Prata2003c},

\begin{equation}
\Omega_{\parallel}(L,d,r)=\frac{1}{2}[1+\frac{1}{\sqrt{1-m^{2}}}\frac {%
(r^{2}-d^{2}-L^{2})}{(r^{2}+d^{2}+L^{2})}]  \label{eq_omega_parallel}
\end{equation}
and from \cite[eqs. 20 and 55]{Prata2003a},

\begin{equation}
\Omega_{\perp}(L,d,r,\alpha)=\cos\alpha\frac{\left\vert L\right\vert }{2d}[%
\frac{1}{\sqrt{1-m^{2}}}-1]~,  \label{eq_omega_ortho}
\end{equation}
where

\begin{equation}
m=2rd/(L^{2}+d^{2}+r^{2})~.  \label{eq_m_def}
\end{equation}
In eq. \ref{eq_omega_ortho} the absolute value of $L$ is used since $L$ is
restricted only by eq. \ref{eq_Lbigger_r} and can thus be negative.

\subsection{{\ Special cases\label{section_special_cases}}}

We emphasize two cases: (i) the axis of symmetry of the disc is parallel to
that of the source ($\beta =0$); and (ii) the center of the disc is located
on the source axis ($R=0$).

\begin{itemize}
\item[(i)] When $\beta =0$, from eqs. \ref{eq_gamma_arccos}, \ref{eq_h_Ld}
and \ref{eq_R_Ld} there results that $\gamma =\alpha $, $h=L$ and $R=d$. The
same results could of course be obtained from \ref{eq_alfa_arccos}, \ref%
{eq_L_hr} and \ref{eq_d_hr}. The restriction imposed by either of the
equivalent eqs. \ref{eq_Lbigger_r} or \ref{eq_hbigger_r} gives $L\geq 0$.
Using eqs. \ref{eq_omega_3}, \ref{eq_omega_parallel} and \ref{eq_m_def} it
is seen that the solid angle is independent from $\alpha $ as expected and

\begin{equation}
\Omega(\beta=0)=\Omega_{\parallel}(L,d,r)
\end{equation}

It was shown in \cite{Prata2003c} that $\Omega_{\parallel}$ is continuous
except for $L\rightarrow0$ since

\begin{equation}
\Omega_{\parallel}(L\rightarrow0)=\left\{ 0~(d>r),1/2~(d=r),1~(d<r)\right\}
~.
\end{equation}

\item[(ii)] When $R=0$, eqs. \ref{eq_alfa_arccos}, \ref{eq_L_hr} and \ref%
{eq_d_hr} give $\alpha =0~,\forall \gamma $, $L=h\cos \beta $ and $d=h\sin
\beta $.

Using eqs. \ref{eq_omega_3}, \ref{eq_omega_parallel}, \ref{eq_omega_ortho}
and \ref{eq_m_def}, a little algebra yields

\begin{equation}
\Omega(R=0)=\frac{\cos\beta~r^{2}}{\sqrt{(r^{2}+h^{2}+2rh\sin\beta
)(r^{2}+h^{2}-2rh\sin\beta)}}~.  \label{eq_omega_R_eq_0}
\end{equation}

$\Omega(R=0)$ is then independent from $\gamma$, as expected.

Eq. \ref{eq_omega_R_eq_0} can be approximated by

\begin{equation}
\Omega(R=0)=\left( \frac{r}{h}\right) ^{2}\cos\beta~\left[ 1-\left( \frac{r}{%
h}\right) ^{2}\cos2\beta+\cdots\right] ~,r\ll h~.  \label{eq_omega_R0_approx}
\end{equation}

In the case where $h=r$, it is straightforward to simplify eq. \ref%
{eq_omega_R_eq_0} to:

\begin{equation}
\Omega(R=0,h=r)=1/2~,\forall\beta~.  \label{eq_omega_R0_eq_r}
\end{equation}
\end{itemize}

\section{Results}

\begin{figure}[h]
\begin{center}
\includegraphics[
height=3.8683cm,
width=6.0934cm
]{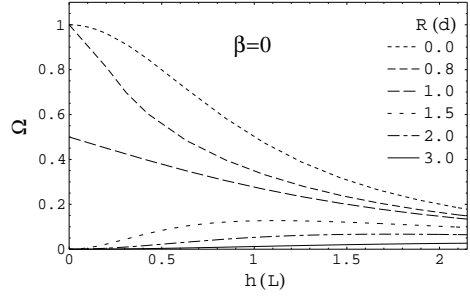}
\end{center}
\caption{Solid angle subtended by a disc of radius 1 with symmetry axis
parallel to that of the source.}
\label{fig4}
\end{figure}

Here we present sample plots of the solid angle. In the following we choose
to work with S' parameters ($\beta,h,R,\gamma$) and consider throughout a
disc of radius $r=1$. We first address the case $\beta=0$, so that $\Omega
=\Omega_{\parallel}$ , $L=h$ and $d=R$. As said before (section \ref%
{section_special_cases}) $\Omega$ is not continuous as $L\rightarrow0$. This
is illustrated in fig. \ref{fig4}\ where $\Omega$ is plotted as a function
of $h(L)$ for different $R(d)$ values.

\begin{figure}[h]
\begin{center}
\includegraphics[
height=3.8661cm,
width=6.0934cm
]{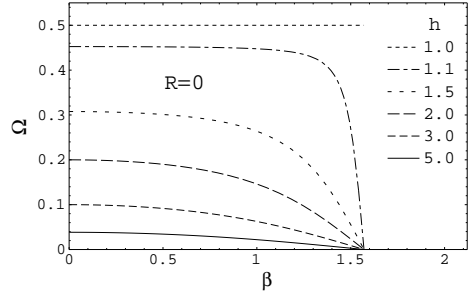}
\end{center}
\caption{Solid angle subtended by a disc of radius 1 and center located on
the source axis.}
\label{fig5}
\end{figure}

\begin{figure}[h]
\begin{center}
\includegraphics[
height=3.7343cm,
width=6.0934cm
]{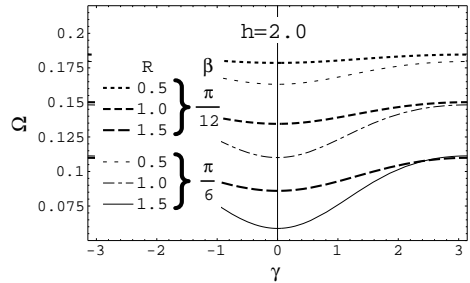}
\end{center}
\caption{Solid angle subtended by a disc of radius 1 for fixed $h=2$.}
\label{fig6}
\end{figure}

\begin{figure}[h]
\begin{center}
\includegraphics[
height=3.8243cm,
width=6.0934cm
]{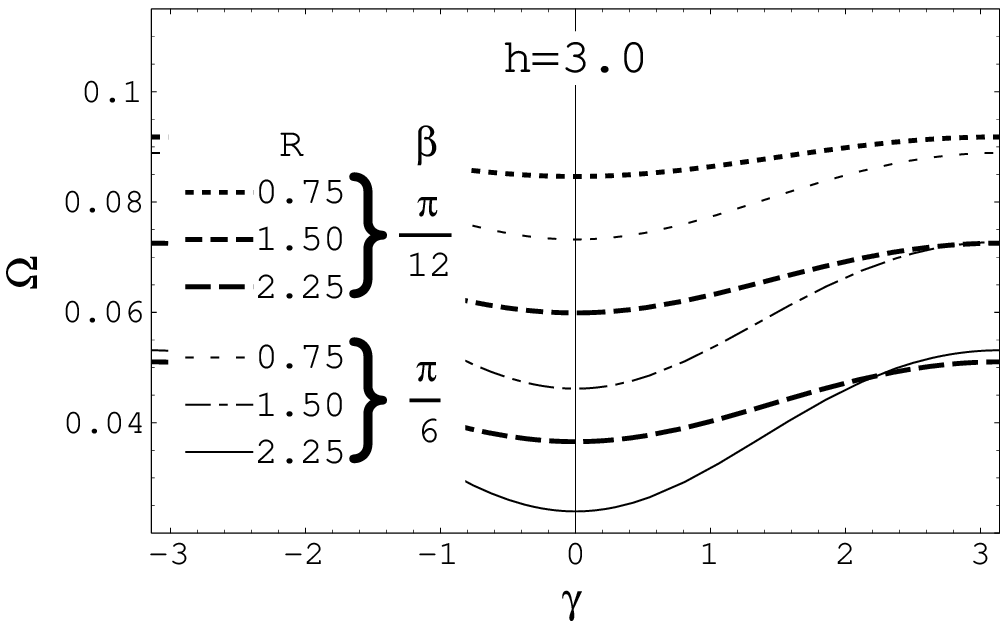}
\end{center}
\caption{Solid angle subtended by a disc of radius 1 for fixed $h=3$.}
\label{fig7}
\end{figure}

In fig. \ref{fig5} the situation where the disc center is on the source axis
($R=0$) is represented. Here, $\Omega$ does not depend on $\gamma$ (section %
\ref{section_special_cases}) and plots of the solid angle as a function of $%
\beta$ are shown for different $h$ values , including the special case $h=r,$
where $\Omega=1/2$, regardless of $\beta$. To avoid restricting the range of
variation of $\beta$ (see eq. \ref{eq_beta_range_h_less_r}), the values of $%
h $ were chosen such that $h\geq r$. In figs. \ref{fig6}, \ref{fig7} and \ref%
{fig8} the effect of offsetting the position of the disc center from the
source axis is shown for three offset values ($R=h/4$, $R=h/2$ and $R=3/4h$)
and for two values of $\beta$ ($\pi/12$, $\pi/6$). As argued before, the
solid angle is an even function of $\gamma$ and it is seen that for the
larger tilting angle ($\beta=\pi/6 $) the dependence on $\gamma$ is enhanced.

\begin{figure}[h]
\begin{center}
\includegraphics[
height=3.8243cm,
width=6.0934cm
]{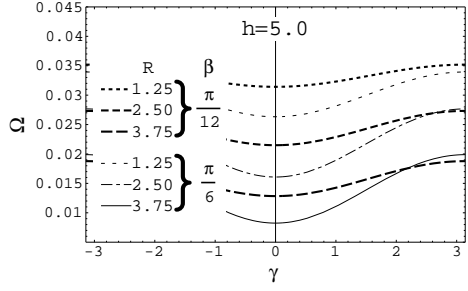}
\end{center}
\caption{Solid angle subtended by a disc of radius 1 for fixed $h=5$.}
\label{fig8}
\end{figure}

\begin{figure}[h]
\begin{center}
\includegraphics[
height=4.8414cm,
width=14.1243cm
]{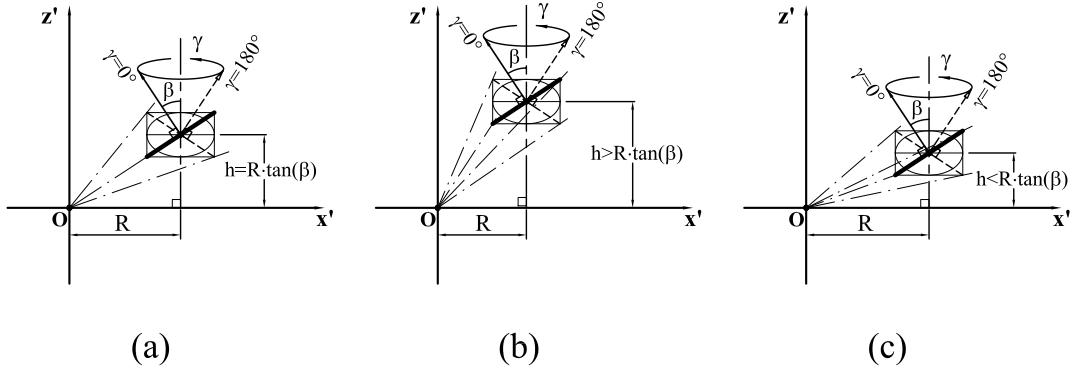}
\end{center}
\caption{The effect of changing $\protect\gamma$ as seen from a side view in
the S' system for: (a) $h=R\cdot tan(\protect\beta)$ , (b) $h>R\cdot tan(%
\protect\beta)$ and (c) $h<R\cdot tan(\protect\beta)$.}
\label{fig9}
\end{figure}

\begin{figure}[h]
\begin{center}
\includegraphics[
height=3.9122cm,
width=6.0912cm
]{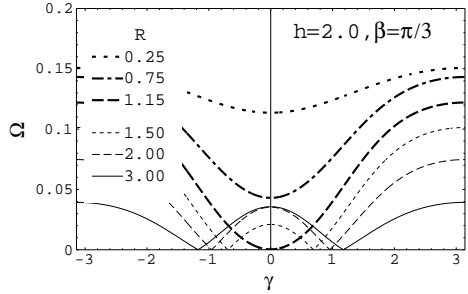}
\end{center}
\caption{Plots of $\Omega$ as a function of $\protect\gamma$ showing a zero
for $h\leq R\cdot tan(\protect\beta)$.}
\label{fig10}
\end{figure}

When plotting $\Omega$ as a function of $\gamma$ while holding all other
parameters constant, there \textit{can} be a zero, i.e. a value of $\gamma$
such that $\Omega(\gamma_{0})=0$. Looking at fig. \ref{fig2} it is clear
that $\Omega=0$ if $L=0$ \textbf{and} $d>r$. Setting $L=0 $ in eq. \ref%
{eq_L_hr} and solving for $\gamma$ gives an equation for the zero,

\begin{equation}
\gamma_{0}=\arccos[h/(R\tan\beta)]~,  \label{eq_gamma0}
\end{equation}

which has no solution for $h/(R\tan\beta)>1$. Thus, this qualitative feature
is dependent on the sign of $h-R\cdot\tan\beta$, which actually
distinguishes the situations where the disc always presents the same face to
the source from those where the face presented depends on the value of $%
\gamma$. This is schematically explained in fig. \ref{fig9} where the effect
of changing $\gamma$ from $0$ to $\pi$ is shown as seen in S' (see also fig.%
\ref{fig1}), when looking along the $y\prime$ axis. If $h-R\cdot\tan\beta=0$
(fig. \ref{fig9}a), the source always 'sees' the lower face of the disc and
for $\gamma=0$ one has $L=0$ and, consequently, $\Omega=0.$ If $h-R\cdot\tan
\beta>0$ (fig. \ref{fig9}b) the source always looks at the lower face of the
disc but $\Omega$ is never zero. Finally, for $h-R\cdot\tan\beta<0$ (fig. %
\ref{fig9}c), it is seen that, as the disc swirls with increasing $\gamma$,
the upper face is first shown (e.g. $\gamma=0$) and then the lower face is
presented (e.g. $\gamma=\pi$), which means that $\Omega=0$ at some point in
between. This behaviour is illustrated in fig. \ref{fig10}, for $h=2$ and $%
\beta=\pi/3$. The zero shows up for $R\geq h/\tan\beta\simeq1.16$.

One should notice that in the preceding discussion it was implicitly assumed
that $d>r$ when $\gamma=\gamma_{0}$. We now proceed to show that eq. \ref%
{eq_hbigger_r}, in the strict form

\begin{equation}
h-r\cdot\sin\beta>0~,  \label{eq_hbigger_r_strict}
\end{equation}
guarantees that $d>r$. Using eq. \ref{eq_gamma0} to eliminate $h$ in eqs. %
\ref{eq_hbigger_r_strict} and \ref{eq_d_hr} gives

\begin{equation}
\cos\gamma\cdot R/\cos\beta>r~
\end{equation}
and

\begin{equation}
d=R/\cos \beta \cdot \sqrt{\cos ^{2}\gamma +\sin ^{2}\gamma \cos ^{2}\beta }%
~.
\end{equation}%
respectively. Since $\sqrt{\cos ^{2}\gamma +\sin ^{2}\gamma \cos ^{2}\beta }%
\geq \cos \gamma $ therefore $d>r$.

\section{Summary and Outlook}

Analytical expressions for the solid angle subtended by a circular disc at a
point cosine source were obtained , under the single restriction that the
disc is located in the half-space illuminated by the source (eq. \ref%
{eq_hbigger_r} or eq. \ref{eq_Lbigger_r}). It was shown (eq. \ref{eq_omega_3}%
) that the solid angle can be decomposed into the combination of two
components ($\Omega _{\parallel}$ and $\Omega_{\perp}$) corresponding to the
situations where the symmetry axis of the disc is parallel (eq. \ref%
{eq_omega_parallel}) or orthogonal (eq. \ref{eq_omega_ortho}) to the source
direction. $\Omega$ can be calculated relatively to two alternative
coordinate systems (S' and S) shown in figs. \ref{fig1} and \ref{fig2}. The
parameters pertaining to each system ($\{\gamma,h,R,(\beta,r)\}$, $%
\{\alpha,L,d,(\beta,r)\}$) are related through eqs. \ref{eq_gamma_arccos}, %
\ref{eq_h_Ld} , \ref{eq_R_Ld}, \ref{eq_alfa_arccos}, \ref{eq_L_hr} and \ref%
{eq_d_hr}.

A similar calculation to that presented here can be performed for the solid
angle defined by a cylindrical detector. A work where we report this latter
result is in preparation.

\begin{ack}
I'm grateful to Jo\~{a}o Prata for rewiewing this manuscript. I would like to thank Professor John H. Hubbell for
providing a copy of the works by A.V. Masket \cite{Mask57}, A.H. Jaffey \cite{Jaff54}
and Hubbell \textit{et al} \cite{Hubb61}.
This work was partially supported by Funda\c{c}\~{a}o para a Ci\^{e}ncia e Tecnologia
(Grant BD/15808/98 - Programa Praxis XXI).
\end{ack}


\begin{thebibliography}{99}
\bibitem{Prata2003a} Prata, M.J., 
Rad. Phys. Chem. \textbf{66} (2003) 387--395. e-print: math-ph/0209065.

\bibitem{Prata2003c} Prata, M.J., 2003. Analytical calculation of the solid
angle defined by a cylindrical detector and a point cosine source with
parallel axes. Accepted Rad. Phys. Chem. (RPC3140). e-print: math-ph/0302003.

\bibitem{Hubb61} Hubbell, J.H., Bach, R.L. and Herbold, R.J., 
J. Research NBS \textbf{65C} (1961) 249--264.

\bibitem{Herm00} Hermann, R.A., A treatise in geometrical optics. (Cambridge
Univ. Press, 1900) 217 (ex. 13). Cited in Ref. \cite{Hubb61}.

\bibitem{Foot15} Foote, P.D., Bull. of NBS \textbf{12} (1915) 583. Cited in
Ref. \cite{Hubb61}.

\bibitem{Knol79} Knoll, G.F., Radiation detection and measurement (John
Wiley \& Sons Ltd, 1979) chap. 3.

\bibitem{Tsou95} Tsoulfanidis, N., Measurement and detection of radiation
(2nd ed, Taylor \& Francis, Bristol, PA, 1995) chap. 8.

\bibitem{Jaff54} Jaffey, A.H., 
Rev. Sci. Instr. \textbf{25} (1954) 349--354.

\bibitem{Mack56} Macklin, P.A., Expression for the solid angle subtended by
a circular disc at a point source in terms of elliptic integrals, included
as a footnote in Ref. \cite{Mask57}.

\bibitem{Mask56} Masket, A.V., Macklin, R.L. and Schmitt, H.W., Tables of
solid angle values and activations, ORNL-2170 (Oak Ridge Nat. Lab., Oak
Ridge, Tenn., 1956)

\bibitem{Mask57} Masket, A.V., 
Rev. Sci. Instr. \textbf{28} (1957) 191--197.

\bibitem{Gard71} Gardner, R.P. and Verghese, K., 
Nucl. Instr. Meth. \textbf{93} (1971) 163--167.

\bibitem{Prata2003b} Prata, M.J., 
Rad. Phys. Chem. \textbf{67} (2003) 599--603. e-print: math-ph/0211061.
\end{thebibliography}
\end{document}